\documentclass[english]{revtex4-1}
\usepackage[LGR,T1]{fontenc}
\usepackage[latin9]{inputenc}
\usepackage{amssymb}
\usepackage{graphicx}
\usepackage{esint}

\makeatletter

\DeclareRobustCommand{\greektext}{%
  \fontencoding{LGR}\selectfont\def\encodingdefault{LGR}}
\DeclareRobustCommand{\textgreek}[1]{\leavevmode{\greektext #1}}
\DeclareFontEncoding{LGR}{}{}
\DeclareTextSymbol{\~}{LGR}{126}

\makeatother

\usepackage{babel}
\begin{document}

\title{Liquid-Gas Phase Transitions and $\mathcal{CK}$ Symmetry in Quantum
Field Theories}

\author{Hiromichi Nishimura}

\email{hnishimura@bnl.gov}

\selectlanguage{english}%

\address{RIKEN BNL Research Center, Brookhaven National Laboratory, Upton
NY 11973 USA}

\author{Michael C. Ogilvie}

\email{mco@physics.wustl.edu}

\selectlanguage{english}%

\author{Kamal Pangeni}

\email{kamalpangeni@wustl.edu}

\selectlanguage{english}%

\address{Department of Physics, Washington University, St. Louis, MO 63130
USA}

\date{12/30/16}

\begin{abstract}
A general field-theoretic framework for the treatment of liquid-gas
phase transitions is developed. Starting from a fundamental four-dimensional
field theory at nonzero temperature and density, an effective three-dimensional
field theory with a sign problem is derived. Although charge
conjugation $\mathcal{C}$ is broken at finite density, 
there remains a symmetry under $\mathcal{CK}$,
where $\mathcal{K}$ is complex conjugation. We consider four models:
relativistic fermions, nonrelativistic fermions, static fermions and
classical particles. The thermodynamic behavior is extracted
from $\mathcal{CK}$-symmetric complex saddle points of the effective
field theory at tree level. The relativistic and static fermions show a liquid-gas
transition, manifesting as a first-order line at low
temperature and high density, terminated by a critical end point.
In the cases of nonrelativistic fermions
and classical particles, we find no first-order liquid-gas transitions
at tree level. 
The mass matrix controlling the behavior of correlation functions
is obtained from fluctuations around the saddle points. Due to the
$\mathcal{CK}$ symmetry of the models, the eigenvalues of the mass
matrix can be complex. This leads to the
existence of disorder lines, which mark the boundaries where the eigenvalues
go from purely real to complex. The regions where the mass matrix
eigenvalues are complex are associated with the critical line. In
the case of static fermions, a powerful duality between particles
and holes allows for the analytic determination of both the critical
line and the disorder lines. Depending on the values of the parameters,
either zero, one or two disorder lines are found. Numerical results
for relativistic fermions give a very similar picture.
\end{abstract}
\maketitle

\section{Introduction}

The problem of determining the phase structure of interacting particles
at nonzero temperature and density is old and important. Modern field-theoretic
approaches are typically susceptible to the sign problem, in which
basic quantities such as the action becomes complex. This problem
is particularly acute in the case of QCD at finite temperature and
density: Lattice simulations have given excellent first-principles
results for many observables of finite-temperature QCD, there has
been less clear success when the chemical potential \textgreek{m}
is nonzero \cite{deForcrand:2010ys,Gupta:2011ma,Aarts:2013bla}. A
central problem is the determination of the phase structure of QCD
at low temperature and density where a critical line with a critical
end point in the Ising, or liquid-gas, universality class is widely
expected.

Here we address the generic problem of liquid-gas phase transitions
from a field theory perspective. The general class of field theories
we will study is the class of $\mathcal{CK}$-symmetric models obtained
from dimensional reduction of a four-dimensional field theory at finite
temperature and density. The simplest case of interest are models
with a single type of particles, interacting via a scalar field $\sigma$
and a vector field $A_{\mu}.$ Both $\sigma$ and $A_{\mu}$ will
be taken to have masses. The potential induced by $\sigma$ will be
attractive, while that caused by the static vector potential $A_{4}$
will be repulsive between particles. The particles of the underlying
theory are integrated out, and after dimensional reduction and redefinition
of fields, we obtain a Lagrangian of the general form
\begin{equation}
L_{3d}=\frac{1}{2}\left(\nabla\phi_{1}\right)^{2}+\frac{1}{2}m_{1}^{2}\phi_{1}^{2}+\frac{1}{2}\left(\nabla\phi_{2}\right)^{2}+\frac{1}{2}m_{2}^{2}\phi_{2}^{2}-F\left(\phi_{1},\phi_{2}\right)
\end{equation}
where $\phi_{1}$ is associated with the attractive force, and $\phi_{2}$
with the repulsive force. The field $\phi_{1}$ is naturally as a
four-dimensional scalar, but $\phi_{2}$ is obtained from the fourth
component of a vector interaction. The function $F$ can be interpreted
as $\beta p\left(\phi_{1},\phi_{2}\right)$, where $\beta$ is the
inverse of the tempurature $T$ and $p$ is a local pressure. In particular,
$p\left(\phi_{1},\phi_{2}\right)$ is the local pressure of the gas
of particles in the grand canonical ensemble in the presence of the
background fields $\phi_{1}$ and $\phi_{2}$. 

The key feature of $L_{3d}$ is that it is not real, but instead satisfies
the $\mathcal{CK}$ symmetry condition 
\begin{equation}
L_{3d}(\phi_{1},\phi_{2})^{*}=L_{3d}(\phi_{1},-\phi_{2}).
\end{equation}
The $\mathcal{C}$ transformation naturally takes $\phi_{2}\rightarrow-\phi_{2}$
as in the case of QED, and the $\phi_{1}$ field is left invariant.
A nonzero chemical potential $\mu$ explicitly breaks $C$ symmetry,
but the antilinear symmetry $\mathcal{CK}$ remains \cite{Meisinger:2012va,Nishimura:2014rxa,Nishimura:2014kla,Nishimura:2015lit}.
$\mathcal{CK}$ symmetry implies that the saddle points of $L_{3d}$
have $\phi_{2}$ purely imaginary; at these saddle points, $L_{3d}$
is real. Analytic continuation of the fields into the complex plane
leads to a resolution of the sign problem at tree level. More generally,
unbroken $\mathcal{CK}$ symmetry implies that the expected value
of $\left\langle \phi_{2}\right\rangle $ must be zero or purely imaginary,
because $\left\langle i\phi_{2}\right\rangle ^{*}=\left\langle i\phi_{2}\right\rangle $. 

The static solutions of the equations of motion take the form
\begin{eqnarray}
m_{1}^{2}\phi_{1} & = & \frac{\partial F}{\partial\phi_{1}}\\
m_{2}^{2}\phi_{2} & = & \frac{\partial F}{\partial\phi_{2}}.
\end{eqnarray}
 The presence of such a $\mathcal{CK}$ symmetry is generic in quantum
field theories at finite density. The mass matrix for $\phi_{1}$
and $\phi_{2}$ is given by

\begin{equation}
\left(\begin{array}{cc}
m_{1}^{2}-\frac{\partial^{2}F}{\partial\phi_{1}^{2}} & -\frac{\partial^{2}F}{\partial\phi_{1}\partial\phi_{2}}\\
-\frac{\partial^{2}F}{\partial\phi_{2}\partial\phi_{1}} & m_{2}^{2}-\frac{\partial^{2}F}{\partial\phi_{2}^{2}}
\end{array}\right).
\end{equation}
The mass matrix is nonhermitian because the off-diagonal elements
are purely imaginary at the saddle point, where $\phi_{2}$ is imaginary.
It does, however, inherit the $\mathcal{CK}$ symmetry of the underlying
model. This in turn implies that the eigenvalues of the mass matrix
are either both real or form a complex conjugate pair \cite{Meisinger:2012va,Nishimura:2014rxa,Nishimura:2014kla,Nishimura:2015lit}
. The boundary of the region where complex conjugate pairs occur is
given by 
\begin{equation}
\left(m_{1}^{2}-\frac{\partial^{2}F}{\partial\phi_{1}^{2}}-m_{2}^{2}+\frac{\partial^{2}F}{\partial\phi_{2}^{2}}\right)^{2}+4\left(\frac{\partial^{2}F}{\partial\phi_{1}\partial\phi_{2}}\right)^{2}=0.
\end{equation}
The occurence of complex conjugate pairs of mass eigenvalues in turn
lead to the damped sinusoidal density oscillations often associated
with the presence of a liquid phase. Thus we see that this formalism
can give a simple understanding of the existence of disorder lines,
which mark a change in the behavior of the potential between particles
from exponential to damped sinusoidal behavior. This behavior cannot
be obtained from field theories with real actions.

One of the hallmarks of typical liquid-gas systems is the existence
of regions in parameter space where density-density correlation functions
exhibit damped oscillatory behavior. From the perspective of quantum
field theory, the appearance of damped oscillatory behavior in correlation
functions is unusual. Such behavior is prohibited in Euclidean quantum
field theories with spectral positivity. The simpest model of the
liquid-gas transition, the Ising model in its binary alloy form, has
spectral positivity and cannot exhibit damped oscillatory behavior
of its correlation functions. Damped oscillatory behavior occurs generally
in the class of models we consider whenever a first-order critical
line occurs.

We will consider four different models, corresponding to four different
choices for the function $F\left(\phi_{1},\phi_{2}\right)$. In the
next section, we will consider the case of relativistic fermions of
mass $m$, with $F$ taken to be
\begin{equation}
F=\int\frac{d^{3}k}{\left(2\pi\right)^{3}}\log\left[1+\exp\left(-\beta\sqrt{k^{2}+\left(m-g\beta^{-1/2}\phi_{1}\right)^{2}}+\beta\mu+i\beta^{1/2}e\phi_{2}\right)\right].
\end{equation}
The coupling constants $g$ and $e$ determine the strength of the
attractive and repulsive forces respectively. There is a natural nonrelativistic
limit, where after redefinition of the chemical potential we have
\begin{equation}
F=\int\frac{d^{3}k}{\left(2\pi\right)^{3}}\log\left[1+\exp\left(-\beta k^{2}/2m+\beta\mu+\beta^{1/2}g\phi_{1}+i\beta^{1/2}e\phi_{2}\right)\right].
\end{equation}
Although the case of relativistic fermions exhibits a first-order
liquid-gas transition, the nonrelativistic reduction does not. The
third section considers the case of static fermions, where
\begin{equation}
F=\frac{1}{v}\log\left[1+\exp\left(-\beta m+\beta\mu+\beta^{1/2}g\phi_{1}+i\beta^{1/2}e\phi_{2}\right)\right]
\end{equation}
where $v$ is a parameter with dimensions of volume. This model also
has a first-order liquid gas transition. However, its low-density
reduction, the classical gas with
\begin{equation}
F=\frac{1}{v}\exp\left(-\beta m+\beta\mu+\beta^{1/2}g\phi_{1}+i\beta^{1/2}e\phi_{2}\right)
\end{equation}
does not have a first-order transition and does not have stable ground
state. On the other hand, its partition function is exactly equivalent
to that of a classical gas, as we show in an appendix. A final section
gives our conclusions.

\section{Relativistic Fermions}

In this section, we study the liquid-gas phase transition in a model
of relativistic fermions. We begin by showing how an effective three-dimensional
field theory can be derived from a fundamental field theory. This
approach is general and can be applied to other models as well, such
as relativistic or nonrelativistic bosons.

\subsection{Derivation from fundamental field theory}

We consider a relativistic fermion interacting with a scalar field
$\sigma$ of mass $m_{\sigma}$ and a vector field $A_{\mu}$ of mass
$m_{A}$. These fields couple to the fermion with coupling $g$ and
$e$ respectively. The Lagrangian for the fermionic part is
\begin{equation}
L_{F}=\bar{\psi}\left[i\gamma\cdot\left(\partial-ieA-\mu\hat{e}_{4}\right)-\left(m-g\sigma\right)\right]\psi
\end{equation}
where $m$ is the fermion mass. The bosonic part is given by
\begin{equation}
L_{B}=\frac{1}{2}\left(\partial\sigma\right)^{2}+\frac{1}{2}m_{\sigma}^{2}\sigma^{2}+\frac{1}{4}F_{\mu\nu}^{2}+\frac{1}{2}m_{A}^{2}A_{\mu}^{2}.
\end{equation}
This can be considered to be a QED version of the PNJL model, that
is, a PNJL model where the gauge symmetry is $U(1)$. We set the boson
masses $m_{\sigma}$ and $m_{A}$ to be constant, although in practical
applications of the formalism they may be temperature dependent. Any
expected value for the boson fields $\sigma$ and $A_{\mu}$ (actually
the associated Polyakov loop) are induced in this model by the effects
of the fermions at finite temperature and density. That said, it is
completely straightforward to include potential terms for $\sigma$
and $A_{4}$ that can produce a more complicated phase structure.
For example, a potential for $\sigma$ could give rise to the analog
of the chiral transition in NJL and PNJL models. In this model, however,
the liquid-gas transition is driven solely by the interactions between
the fermions. 

The finite-temperature one-loop effective potential in the presence
of constant fields $\sigma$ and $A_{4}$ takes the form

\begin{equation}
S_{eff}=\int d^{d}x\left[V_{\phi}+V_{A}+V_{FT}\right]
\end{equation}
where the temperature-dependent part of the one-loop fermionic contribution
to the effective potential $V_{FT}$ is given by
\begin{equation}
V_{FT}=-\frac{1}{\beta}\int\frac{d^{3}k}{\left(2\pi\right)^{3}}\log\left[1+\exp\left(-\beta\sqrt{k^{2}+\left(m-g\sigma\right)^{2}}+\beta\mu+i\beta eA_{4}\right)\right]
\end{equation}
and $V_{\sigma}$ and $V_{A}$ are quadratic. Although there are also
contributions from thermal excitations of $\sigma$ and $A_{\mu}$,
we ignore them here because they do not affect the phase structure.
For simplicity, we generally assume that $\mu$ is sufficently large
that the antiparticle contribution can be suppressed. However, in
the case of relativistic fermions, antiparticle effects must be included
to obtain the correct phase structure near $\mu=0$ , so in this case
antiparticle effects are included. For comparison with the other models,
we do no denote these effect explicitly here. The potential $V_{FT}$
is nothing but the negative of the pressure of a relativistic fermion
moving in the constant background provided by $\sigma$ and $A_{4}$.
For slowly varying fields $\sigma$ and $A_{\mu}$, $V_{FT}$ represents
the lowest order fermionic contribution in a derivative expansion
of the effective action \cite{Coleman:1973jx}. We have also assumed
that the spatial part of the vector field can be neglected, so that
only the timelike component of $A_{\mu}$ need be included.

We dimensionally reduce to a three-dimensional effective theory, yielding
an effective Lagrangian $L_{3d}$ of the form
\[
L_{3d}=\beta L_{B}+\beta V_{FT}.
\]
 We define new variables $\phi_{1}=\beta^{1/2}\sigma$ and $\phi_{2}=\beta^{1/2}A_{4}$
so as to maintain the canonical kinetic terms and also set $m_{1}=m_{\sigma}$
and $m_{2}=m_{A}$, giving
\begin{eqnarray}
L_{3d} & = & \frac{1}{2}\left(\nabla\phi_{1}\right)^{2}+\frac{1}{2}m_{1}^{2}\phi_{1}^{2}+\frac{1}{2}\left(\nabla\phi_{2}\right)+\frac{1}{2}m_{2}^{2}\phi_{2}^{2}\nonumber\\
 &  & -\int\frac{d^{3}k}{\left(2\pi\right)^{3}}\log\left[1+\exp\left(-\beta\sqrt{k^{2}+\left(m-g\beta^{-1/2}\phi_{1}\right)^{2}}+\beta\mu+i\beta^{1/2}e\phi_{2}\right)\right].
\end{eqnarray}
The last term in the expression is the definition of $F(\phi_{1},\phi_{2})$
for this model.

\subsection{Phase structure and disorder lines}

The phase structure of the model is obtained from the static solutions
of the equations of motion:
\begin{eqnarray}
m_{1}^{2}\phi_{1} & = & \frac{\partial F}{\partial\phi_{1}}\\
m_{2}^{2}\phi_{2} & = & \frac{\partial F}{\partial\phi_{2}}
\end{eqnarray}
 while the presence of disorder lines is determined from the mass
matrix
\begin{equation}
\left(\begin{array}{cc}
m_{1}^{2}-\frac{\partial^{2}F}{\partial\phi_{1}^{2}} & -\frac{\partial^{2}F}{\partial\phi_{1}\partial\phi_{2}}\\
-\frac{\partial^{2}F}{\partial\phi_{2}\partial\phi_{1}} & m_{2}^{2}-\frac{\partial^{2}F}{\partial\phi_{2}^{2}}
\end{array}\right).
\end{equation}
In addition to $T$ and $\mu,$ all of the models we consider have
five parameters: $m$, $m_{1}$, $m_{2}$, $e$and $g$. This is a
very large parameter space to explore. In general, the potential takes
the form $F(\phi_{1},\phi_{2})\rightarrow F(g\phi_{1},e\phi_{2})$.
In terms of the rescaled fields $\tilde{\phi}_{1}=g\phi_{1}$ and
$\tilde{\phi}_{2}=e\phi_{2}$, we can write the eqution of motion
as 
\begin{eqnarray}
\frac{1}{\kappa_{1}}\tilde{\phi}_{1} & = & \frac{\partial F}{\partial\tilde{\phi}_{1}}\\
\frac{1}{\kappa_{2}}\tilde{\phi}_{2} & = & \frac{\partial F}{\partial\tilde{\phi}_{2}}
\end{eqnarray}
where $\kappa_{1}=g^{2}/m_{1}^{2}$ and $\kappa_{2}=e^{2}/m_{2}^{2}$.
It is then clear that in addition to $T$ and $\mu,$ only three parameters,
$\kappa_{1}$, $\kappa_{2}$ and $m$ determine the solution as well
as the location of the critical line if there is one. On the other
hand, the equation for the disorder line becomes 
\begin{equation}
\left[\left(\frac{1}{\kappa_{1}}-\frac{\partial^{2}F}{\partial\tilde{\phi}_{1}^{2}}\right)-\frac{e^{2}}{g^{2}}\left(\frac{1}{\kappa_{2}}-\frac{\partial^{2}F}{\partial\tilde{\phi}_{2}^{2}}\right)\right]^{2}+4\frac{e^{2}}{g^{2}}\left(\frac{\partial^{2}F}{\partial\tilde{\phi}_{1}\partial\tilde{\phi}_{2}}\right)^{2}=0
\end{equation}
so the disorder line depends on the additional parameter of $e/g$.
Because we are interested in conventional liquid gas transitions,
we will choose the fermion mass $m$ to be substantially heavier than
the masses $m_{1}$ and $m_{2}$. In all the models we consider, we
set $m=20$, $\kappa_{1}=1$, and $e=0.3$, and we then vary the value
of $\kappa_{2}$ to see the change of phase diagramas, as well as
the value of $g=m_{1}$ to observe the difference in disorder lines. 

\begin{figure}
\includegraphics[width=6.5in]{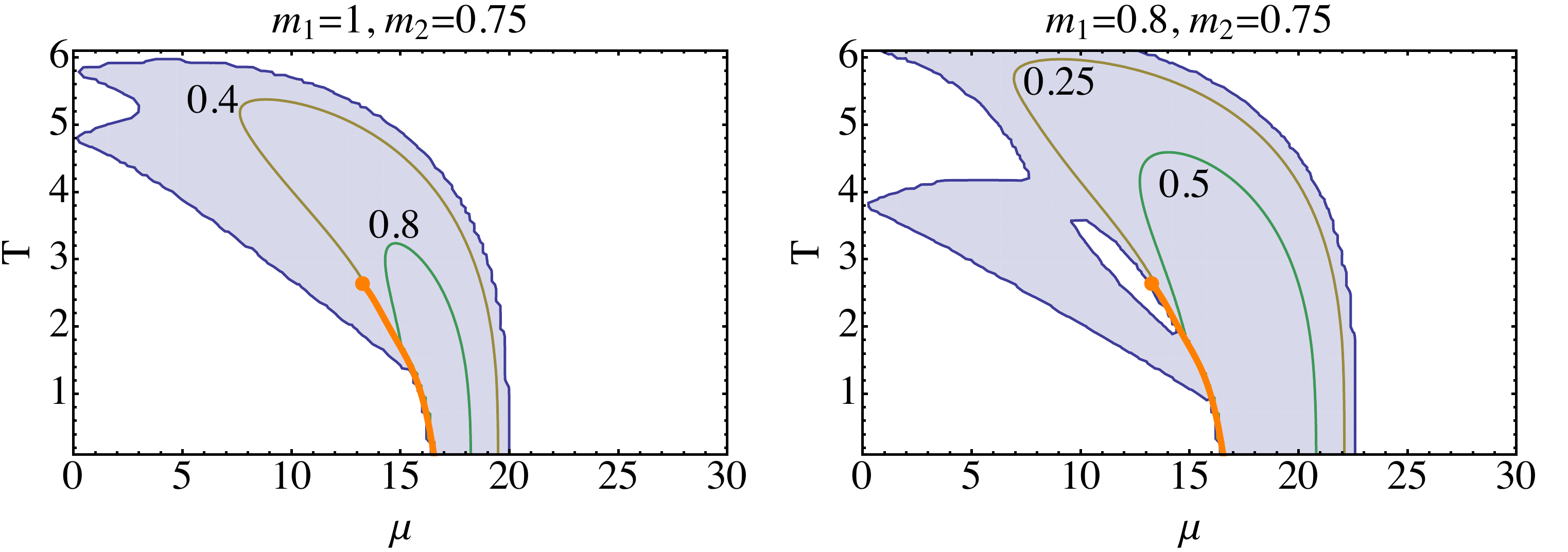}

\caption{\label{fig:RF_LargeKappa}Phase diagrams for relativistic fermions
for $m=20$ and $m_{2}=0.75$ with $e=0.3$. In the first graph $g_{1}=m_{1}=1$,
while in the second $g_{1}=m_{1}=0.8$. The shaded region indicates
where the mass matrix eigenvalues form complex conjugate pairs, and
the contour lines refer to the imaginary parts of the mass matrix
eigenvalues. The boundary of the shaded region defines the disorder
line in the phase diagram. Note the appearance of a second disorder
line inside the first in the second graph. The thick line shows a
first-order line emerging from the $T=0$ axis and terminating in
a critical end point. }
\end{figure}

Generally speaking, the liquid-gas transition will occur for low temperatures
and $\mu\lesssim m$. The left-hand graph of Figure \ref{fig:RF_LargeKappa}
shows the phase diagram for $m=20$, $m_{1}=1$ and $m_{2}=0.75$.
The couplings are given by $e=0.3$ and $g=1$. The shaded region
indicates where the mass matrix eigenvalues form complex conjugate
pairs, and the contour lines refer to the imaginary parts of the mass
matrix eigenvalues. The boundary of the shaded region defines the
disorder line in the phase diagram. The thick line shows a first-order
line emerging from the $T=0$ axis and terminating in a critical end
point. The disorder line has a somewhat surprising shape; we will
return to this point later. The graph on the right-hand side of the
figure shows what happens if $m_{1}$ and $g$ are decreased to $0.8$.
The phase structure is essentially unchanged, and the old disorder
line has changed only slightly. However, a new disorder line boundary
has opened up near the critical end point, inside the region where
complex mass matrix eigenvalues were previously found.

\begin{figure}
\includegraphics[width=6.5in]{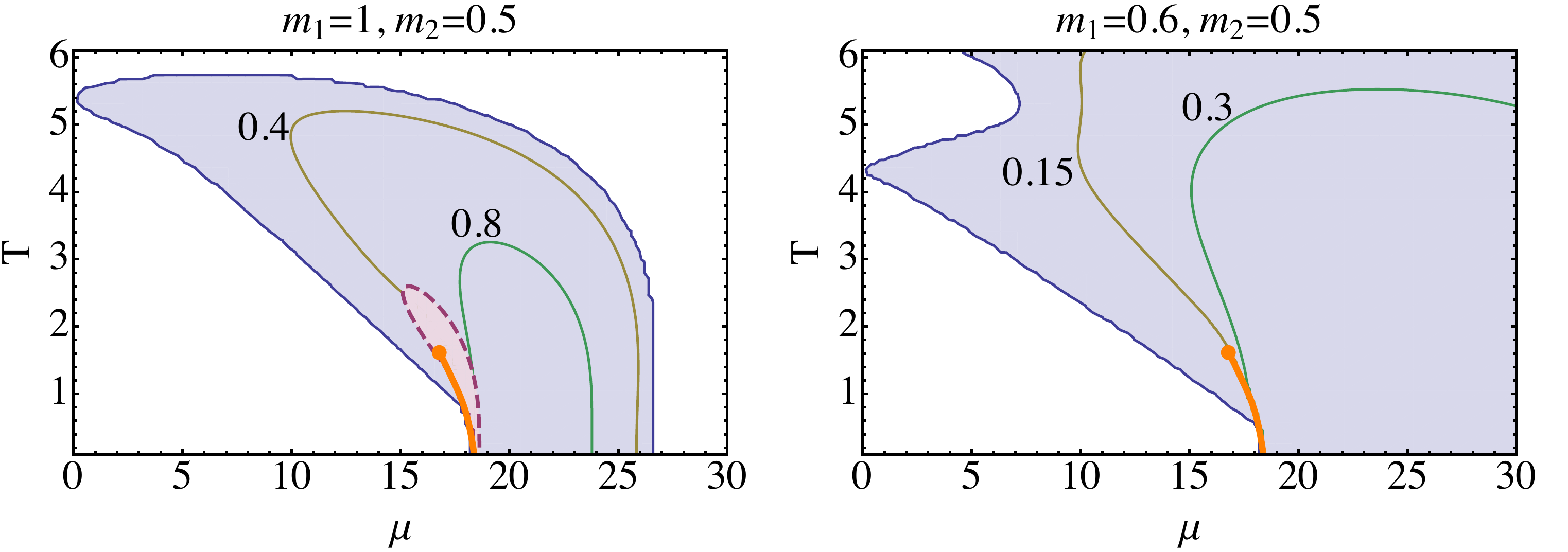}

\caption{\label{fig:RF_SmallKappa}Phase diagrams for relativistic fermions
for $m=20$ and $m_{2}=0.5$ with $e=0.3$. In the first graph $g_{1}=m_{1}=1$,
while in the second $g_{1}=m_{1}=0.6$. The shaded region indicates
where the mass matrix eigenvalues form complex conjugate pairs, and
the contour lines refer to the imaginary parts of the mass matrix
eigenvalues. The boundary of the shaded region defines the disorder
line in the phase diagram. In the first graph, the dashed line near
the critical end point is the boundary of the regions where the real
parts of the mass matrix eigenvalues are negative. The thick line
shows a first-order line emerging from the $T=0$ axis and terminating
in a critical end point. }

\end{figure}

In Figure \ref{fig:RF_SmallKappa}, we show a second pair of phase
diagrams. The graph on the left-hand side has $m_{1}=g=1$ and $m_{2}=0.5$.
As before, $e=0.3$ and $m=20$. The end point of the critical line
is at a lower value of $T$ and slightly shifted to the right, but
is otherwise similar to the previous graphs. However, when we examine
the eigenvalues of the mass matrix, we see something new: the real
part of conjugate pair of mass matrix eigenvalues becomes negative
in a region near the critical end point. This is not neccesarily unphysical
behavior. The mass matrix is the matrix of squared masses, which are
in general complex. A sufficiently large phase in the complex mass
will lead to a squared mass eigenvalue with a negative real part.
The boundary of this region is denoted in the figure by a dashed line.
We have checked carefully for alternative possibilities and have concluded
that this is likely to represent the correct phase structure of the
model. We will return to this point in our conclusions after examining
results from the other models. The graph on the right-hand side of
the figure has $m_{1}=g=0.6$ and again has $m_{2}=0.5$, $e=0.3$
and $m=20$. The lower values of $g$ and $m_{1}$ eliminate the region
where the real part of the conjugate pair of mass matrix eigenvalues
becomes negative. The shaded region becomes larger, but the values
of the imaginary parts become smaller.

\subsection{Nonrelativistic Fermions}

The case of nonrelativistic fermions can be obtained straightforwardly
from the relativistic case, yielding the effective Lagrangian
\begin{eqnarray}
L_{3d} & = & \frac{1}{2}\left(\nabla\phi_{1}\right)^{2}+\frac{1}{2}m_{1}^{2}\phi_{1}^{2}+\frac{1}{2}\left(\nabla\phi_{2}\right)+\frac{1}{2}m_{2}^{2}\phi_{2}^{2}\nonumber\\
 &  & -\int\frac{d^{3}k}{\left(2\pi\right)^{3}}\log\left[1+\exp\left(-\beta k^{2}/2m+\beta\mu-\beta m+\beta^{1/2}g\phi_{1}+i\beta^{1/2}e\phi_{2}\right)\right].
\end{eqnarray}
An important simplification occurs because the fields $\phi_{1}$
and $\phi_{2}$ appear in $F$ only as the combination $\Phi=\beta^{1/2}g\phi_{1}+i\beta^{1/2}e\phi_{2}$.
The equations of motion for static solutions become
\begin{equation}
\phi_{1}=\frac{\beta^{1/2}g}{m_{1}^{2}}\frac{\partial F}{\partial\Phi}
\end{equation}
\begin{equation}
\phi_{2}=i\frac{\beta^{1/2}e}{m_{2}^{2}}\frac{\partial F}{\partial\Phi}.
\end{equation}
Combining these equations , we obtain
\begin{equation}
\Phi=\left(\frac{\beta g^{2}}{m_{1}^{2}}-\frac{\beta e^{2}}{m_{2}^{2}}\right)\frac{\partial F}{\partial\Phi}.
\end{equation}
Defining
\begin{equation}
\kappa=\kappa_{1}-\kappa_{2}=\frac{g^{2}}{m_{1}^{2}}-\frac{e^{2}}{m_{2}^{2}}
\end{equation}
 we see that the phase diagram is determined by the single equation
\begin{equation}
\Phi=\beta\kappa\frac{\partial F}{\partial\Phi}
\end{equation}
 with the four parameters $m_{1}$, $m_{2}$, $e$ and $g$ collapsing
into a single parameter $\kappa$. The solutions of this equation
are extrema of the synthetic potential 
\begin{equation}
U=\frac{1}{2}\Phi^{2}-\beta\kappa F\left(\Phi\right).
\end{equation}
This simplification also holds for static and classical particles
as well. 

Unlike the case of relativistic fermions, we find no liquid-gas transition
in the case of nonrelativistic fermions. This can be very simply from
the behavior of the potential $U$, which always has a single minimum.

\section{Static Fermions}

In this section, we study the behavior of static continuum fermions,
which have no kinetic energy. The potential is given by
\begin{equation}
V_{static}=-\frac{1}{\beta v}\log\left[1+\exp\left(-\beta m+\beta g\sigma+\beta\mu+i\beta eA_{4}\right)\right]
\end{equation}
where $v$ should be thought of as some volume associated with the
particle. For lattice models, this is a natural limit at nonzero temperature
where very heavy particles are fixed on a spatial lattice site. For
continuum field theories, there is no systematic approximation which
yields static fermions as a natural limit. Aside from the connection
with lattice gauge theory, there are nevertheless good reasons to
consider this model. The lattice form of this model was studied by
Park and Fisher as a tool for demonstrating that the repulsive-core
phase transition at negative $z\equiv\exp\beta\left(\mu-m\right)$
is in the $i\phi^{3}$ universality class \cite{ParkFisher1999}.
The continuum model plays a similar role, illustrating both a liquid-gas
transition in the usual Ising universality class and a repulsive-core
transition in the $i\phi^{3}$ universality class for $z<0$. The
model is also interesting because it has an exact particle-hole symmetry
that allows us to determine analytically the location of the critical
line as well as some of the other key features of the model. This
model reduces to the classical model in the limit where $z\exp\Phi\ll1$.
In that case the parameter $v$ can be identified as $\lambda_{T}^{3}=(2\pi/mT)^{3/2}$,
but that identification is special to the low-density limit.

The dimensionally reduced effective Lagrangian $L_{3d}$ has the form

\begin{equation}
L_{3d}=\frac{1}{2}\left(\nabla\phi_{1}\right)^{2}+\frac{1}{2}m_{1}^{2}\phi_{1}^{2}+\frac{1}{2}\left(\nabla\phi_{2}\right)^{2}+\frac{1}{2}m_{2}^{2}\phi_{2}^{2}-\frac{1}{v}\log\left[1+\exp\left(-\beta m+\beta^{1/2}g\phi_{1}+\beta\mu+i\beta^{1/2}e\phi_{2}\right)\right].
\end{equation}
As was the case with nonrelativistic fermions, the crucial simplifying
feature of this model, is that $F$ depends on $\phi_{1}$ and $\phi_{2}$
only through $\Phi=\beta^{1/2}g\phi_{1}+i\beta^{1/2}e\phi_{2}$ .
The static equations of motion reduce to
\begin{equation}
\Phi=\beta\kappa\frac{\partial F}{\partial\Phi}
\end{equation}
which in this case can be written as
\begin{equation}
\Phi=\left(\frac{\beta\kappa}{v}\right)\frac{\partial}{\partial\Phi}\log\left[1+z\exp\left(\Phi\right)\right].
\end{equation}
The corresponding potential $U$ takes the form
\begin{equation}
U=\frac{1}{2}\Phi^{2}-\kappa y\log\left[1+z\exp\left(\Phi\right)\right]
\end{equation}
where we have introduced for convenience $y=\beta/v$.

This model has a conventional liquid-gas transtion for $\kappa y>0$
and $z>0$. We can locate the critical point of this model analytically.
The second derivative of the potential $d^{2}U/d\Phi^{2}$ has two
inflection points when $\kappa y>4$; hence the potential $U$ itself
has two minima for $\kappa y>4$. At the critical end point, the minimum
of the potential coalesces with the two inflection points and one
finds the critical end point at $\left(\kappa y=4,z=e^{-2}\right)$.
Note that for $\kappa<0$, there is a phase transition for $z<0$;
this transition is in the $i\phi^{3}$ universality class as discussed
by Park and Fisher\cite{ParkFisher1999}. As they show, the phase
structure may also be understood graphically. The equation $\partial U/\partial\Phi=0$
may easily be written in the form
\begin{equation}
\Phi\exp\left(-\Phi\right)=\kappa yz-z\Phi
\end{equation}
and the solutions found from the intersection of the left- and right-hand
sides.

\subsection{Particle-Hole Duality}

Analytic information about the phase structure may be obtained from
an exact dualtity argument \cite{Nishimura:2015lit} that exchanges
particle and holes. We can rewrite $U$ as
\begin{equation}
U=\frac{1}{2}\Phi^{2}-\kappa y\log\left[1+z^{-1}e^{-\Phi}\right]-\kappa y\Phi-\kappa y\log z
\end{equation}
or
\begin{equation}
U=\frac{1}{2}\left(\Phi-\kappa y\right)^{2}-\kappa y\log\left[1+z^{-1}e^{-\Phi}\right]-\kappa y\log z-\frac{1}{2}\left(\kappa y\right)^{2}.
\end{equation}
After shifting $\Phi\rightarrow\Phi'=-\Phi+\kappa y$, we have
\begin{equation}
U=\frac{1}{2}\Phi^{2}-\kappa y\log\left[1+z^{-1}e^{\Phi-\kappa y}\right]-\kappa y\log z-\frac{1}{2}\left(\kappa y\right)^{2}
\end{equation}
so the phase structure as revealed by $\Phi$ is invariant under
\begin{equation}
z\rightarrow z'=z^{-1}e^{-\kappa y}
\end{equation}
\begin{equation}
\Phi\rightarrow\Phi'=-\Phi+\kappa y
\end{equation}
These results can be extended to the potential $V,$ where the duality
transformation acts on $\phi_{1}$ and $\phi_{2}$ as
\begin{eqnarray}
\phi_{1} & \rightarrow & \phi_{1}^{'}=-\phi_{1}+\frac{\beta^{1/2}g}{vm_{1}^{2}}\\
\phi_{2} & \rightarrow & \phi_{2}^{'}=-\phi_{2}+\frac{i\beta^{1/2}e}{vm_{2}^{2}}
\end{eqnarray}
consistent with the duality transformation of $\Phi$.

The critical line must map into itself under this transformation and
thus must form part of the curve
\begin{equation}
z=e^{-\kappa y/2}.
\end{equation}
This is
\begin{equation}
\mu=m-\frac{\kappa}{2v}=m-\frac{1}{2v}\left(\frac{g^{2}}{m_{1}^{2}}-\frac{e^{2}}{m_{2}^{2}}\right).
\end{equation}
The critical end point at $\left(\kappa y=4,z=e^{-2}\right)$ lies
on the critical line and maps onto itself under duality. In more physial
units, we have for the critical end point $T_{cep}=\kappa/4v$ and
$\mu_{cep}=m-2T_{cep}$. Along the critical line, the jump in $\Phi$
is given by
\begin{equation}
\Delta\Phi=\Phi-\Phi'=2\Phi-\kappa y.
\end{equation}
This is zero at the critical end point, which must occur when $\Phi=\kappa y/2$,
consistent with the location of the critical end point.

The disorder lines associated with the mass matrix
\begin{equation}
\left(\begin{array}{cc}
m_{1}^{2}-\frac{\partial^{2}F}{\partial\phi_{1}^{2}} & -\frac{\partial^{2}F}{\partial\phi_{1}\partial\phi_{2}}\\
-\frac{\partial^{2}F}{\partial\phi_{2}\partial\phi_{1}} & m_{2}^{2}-\frac{\partial^{2}F}{\partial\phi_{2}^{2}}
\end{array}\right)
\end{equation}
occur when the two eigenvalues are degenerate. Using the fact is a
function of $\Phi$, we arrive after some algebra at the condition
\begin{equation}
m_{1}^{2}-m_{2}^{2}-\beta\left(g\pm e\right)^{2}\frac{\partial^{2}F}{\partial\Phi^{2}}=0.
\end{equation}
This equation for the disorder lines holds whenever $F$ is a function
only of $\Phi$ rather than $\phi_{1}$ and $\phi_{2}$ separately;
this includes the cases of the nonrelativistic fermionic gas and the
classical gas as well as the case of static fermions. For static fermions,
this equation may be written as
\begin{equation}
m_{1}^{2}-m_{2}^{2}-\beta\left(g\pm e\right)^{2}\frac{1}{v}\frac{z\exp\Phi}{\left(1+z\exp\Phi\right)^{2}}=0.
\end{equation}
Because the first term in the sum is always postive and the second
and third terms are always negative, we must have $m_{1}\ge m_{2}$
for disorder lines to appear. The appearance of a factor $\left(g\pm e\right)^{2}$
in the third term makes possible the appearance of two distinct disorder
lines, defining two distinct boundaries for the region where the mass
matrix eigenvalues have imaginary parts. The negative contribution
of the third term is larger in magnitude for the combination $g+e$;
this indicates that it is possible to have zero, one or two disorder
lines.

\subsection{Phase structure and disorder lines}

\begin{figure}
\includegraphics[width=6.5in]{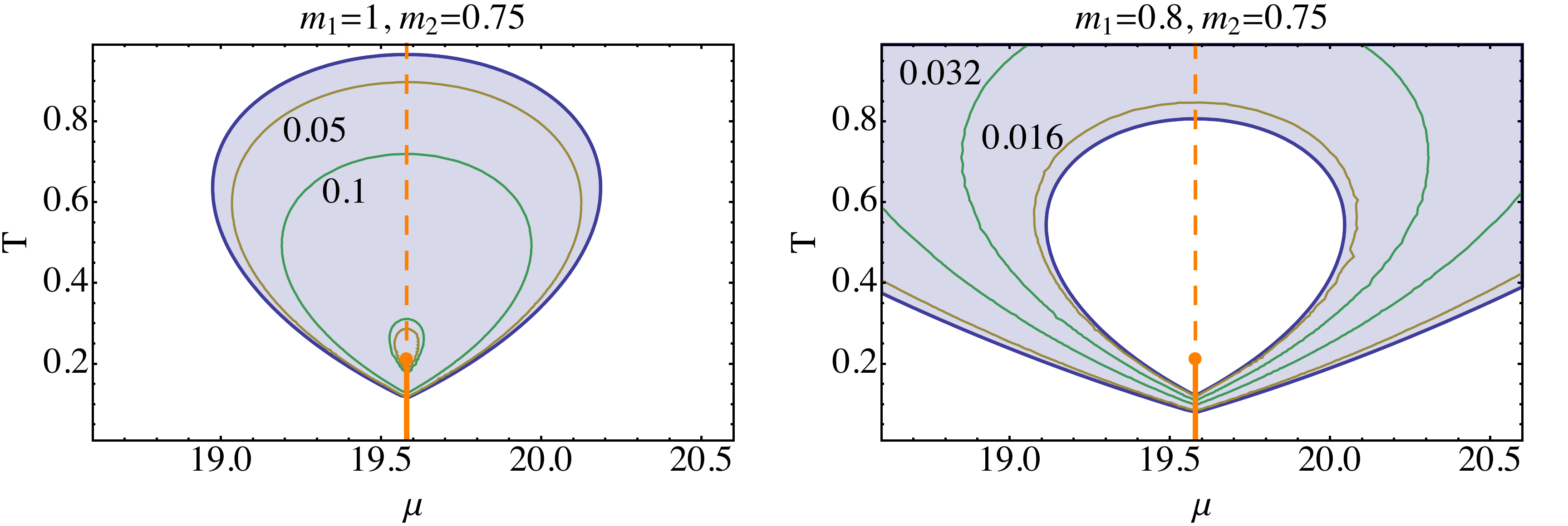}

\caption{\label{fig:SF_LargeKappa}Phase diagrams for static fermions for $m=20$
and $m_{2}=0.75$ with $e=0.3$. In the first graph $g_{1}=m_{1}=1$,
while in the second $g_{1}=m_{1}=0.8$. The shaded region indicates
where the mass matrix eigenvalues form complex conjugate pairs, and
the contour lines refer to the imaginary parts of the mass matrix
eigenvalues. The boundary of the shaded region defines the disorder
line in the phase diagram. Note the appearance of a second disorder
line inside the first in the second graph. The thick line shows a
first-order line emerging from the $T=0$ axis and terminating in
a critical end point. The dashed vertical line emerging from the critical
end point is the line of particle-hole duality.}

\end{figure}

As we did in the case of relativistic fermions, we will choose the
fermion mass $m$ to be substantially heavier than the masses $m_{1}$
and $m_{2}$, with $m=20$. The left-hand graph of Figure \ref{fig:SF_LargeKappa}
shows the phase diagram for $m=20$, $m_{1}=1$ and $m_{2}=0.75$.
The couplings are given by $e=0.3$ and $g=1$. These are exactly
the same values as those used for the first graph in Figure \ref{fig:RF_LargeKappa}.
We set $v=1$ throughout. The vertical line is the line of particle-hole
self-duality; the lower portion of this line is a line of first-order
phase transitions, terminated by a critical end point. The shaded
region again indicates where the mass matrix eigenvalues form complex
conjugate pairs, and the contour lines refer to the imaginary parts
of the mass matrix eigenvalues. The reflection symmetry of the diagram
about the self-dual line is due to particle-hole duality. With these
parameters, we see that there is a single disorder line. The graph
on the right-hand side of the figure shows what happens if $m_{1}$
and $g$ are decreased to $0.8$ while $m_{2}$ is set to $0.75$.
As was the case with relativistic fermions, the phase structure is
essentially unchanged, but a new disorder line boundary has opened
up around the critical end point, inside the region where complex
mass matrix eigenvalues were previously found.

\begin{figure}
\includegraphics[width=6.5in]{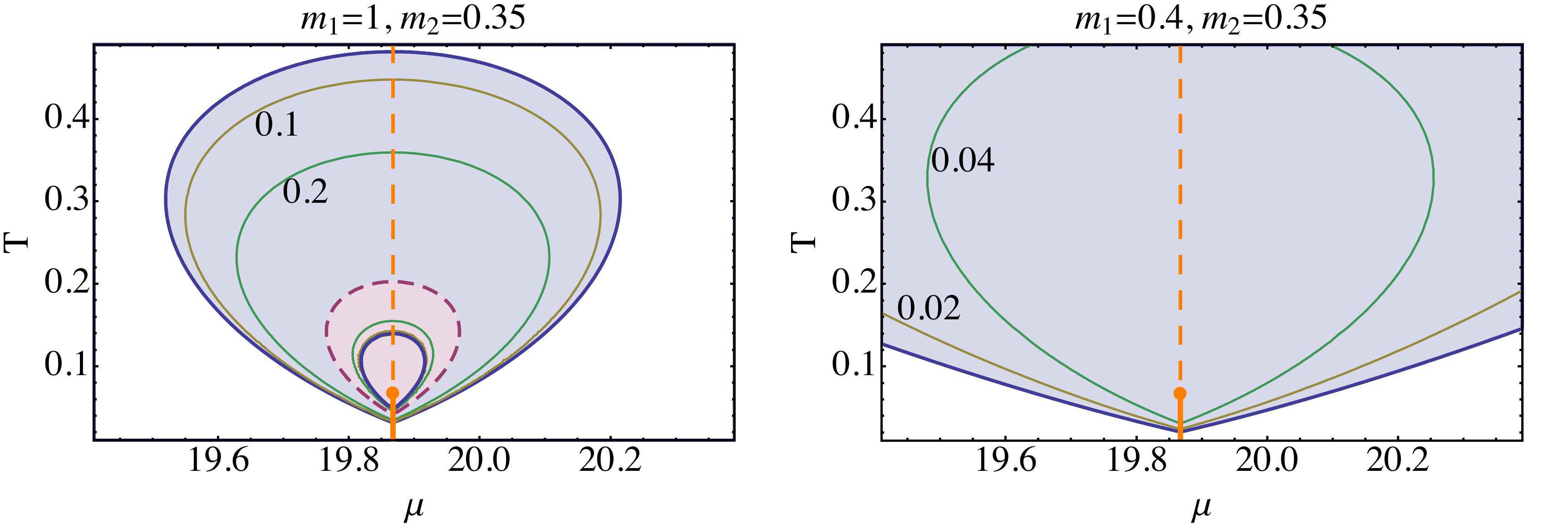}

\caption{\label{fig:SF_SmallKappa}Phase diagrams for static fermions for $m=20$
and $m_{2}=0.35$ with $e=0.3$. In the first graph $g_{1}=m_{1}=1$,
while in the second $g_{1}=m_{1}=0.4$. The shaded region indicates
where the mass matrix eigenvalues form complex conjugate pairs, and
the contour lines refer to the imaginary parts of the mass matrix
eigenvalues. The boundary of the shaded region defines the disorder
line in the phase diagram. In the first graph, the dashed line near
the critical end point is the boundary of the regions where the real
parts of the mass matrix eigenvalues are negative. The thick line
shows a first-order line emerging from the $T=0$ axis and terminating
in a critical end point. The dashed vertical line emerging from the
critical end point is the line of particle-hole duality.}

\end{figure}

In Figure \ref{fig:SF_SmallKappa}, we show a second pair of phase
diagrams. The graph on the left-hand side has $m_{1}=g=1$ and $m_{2}=0.35$.
As was the case with relativistic fermions, we find a small region
around the critical end point where real parts of the eigenvalues
of the mass matrix become negative. The boundary of this region is
again denoted by a dashed line. The graph on the right-hand side of
the figure has $m_{1}=g=0.4$ and $m_{2}=0.35$. The lower values
of $g$ and $m_{1}$ again eliminate the region where the real part
of the conjugate pair of mass matrix eigenvalues becomes negative.
The shaded region becomes larger, but the values of the imaginary
parts become smaller. This behavior is again similar to what we found
for relativistic fermions.

\subsection{Classical Particles}

It is easy to obtain the effective field theory assoiciated with the
classical gas as a limit of the static fermion case. As mentioned
previously, the correct behavior is obtained when $z\exp\Phi\ll1$.
The potential $U$ can be written as 
\begin{equation}
U=\frac{1}{2}\Phi^{2}-\kappa yz\exp\left(\Phi\right).
\end{equation}
The equation $\partial U/\partial\Phi=0$ now becomes
\begin{equation}
\Phi\exp\left(-\Phi\right)=\kappa yz.
\end{equation}
This equation is solved by the Lambert function $W$: $\Phi=-W\left(-\kappa yz\right).$
More intuitively, it may be solved graphically by plotting $\Phi e^{-\Phi}$,
which must be $\kappa zy.$ It is easy to see that there is no real
solution for $\kappa yz>e^{-1}$, two solutions for $0<\kappa yz<e^{-1}$
and one solution for $\kappa yz<0$. Nowhere do we obtain the three
solutions that would be expected with a standard first-order phase
transition: two local minima separated by a local maximum. We can
also visualize this result by noting that $U$ is unbounded from below
for $\kappa yz>0$, with a local maximum and local minimum when $0<\kappa yz<e^{-1}$
, and no extrema for $\kappa yz>e^{-1}$. When $\kappa yz=e^{-1}$,
there is a single static solution at $\Phi=1$. It is easy to confirm
that the mass matrix has a zero eigenvalue at this point, but it is
not a conventional critical point; because it is unstable at cubic
order, it is more like a spinodal point, where a metastable solution
becomes unstable. It has been known for some time that a straightforward
application of mean field theory to the classical liquid-gas system
is insufficient to recover the critical behavior \cite{HUBBARD1972245,Brillantov1998};
it is therefore perhaps unsurprising that tree-level in an equivalent
field theoretic approach is also insufficient.

\section{Conclusions}

We have developed a framework for deriving and analyzing field-theoretic
models of liquid-gas transtitions and applied the formalism to some
important models. In this framework, it is necessary to have two or
more fields to include the effects of both attractive and repulsive
potentials. The presence of a repulsive potential at nonzero $\mu$
gives rise to a sign problem in this class of field theories. Although
charge conjugation symmetry is explicitly broken when $\mu\ne0$,
the symmetry $\mathcal{CK}$ is unbroken, with profound consequences.
One consequence of the $\mathcal{CK}$ symmetry is that there are
regions of the phase diagram where masses have imaginary parts, givin
rise to damped oscillatory behavior in correlation functions. The
border of these regions are disorder lines. This behavior cannot occur
in conventional field theories without sign problems, as a consequence
of spectral positivity.

We have found two models, relativistic fermions and static fermions,
that have conventional liquid-gas transitions at tree level. In contrast,
the field theories associated with nonrelativistic fermions and classical
particles do not have liquid-gas transtions at tree level. The case
of static fermions has proven to be very tractable due to the exact
particle-hole duality found there. As in our previous work on PNJL-type
models of QCD \cite{Nishimura:2014rxa,Nishimura:2014kla}, the critical
line of the liquid-gas transition is generally found in the phase
diagram near any disorder lines present, although there does not seem
to be any simple universal rule. The occurence of zero, one or two
disorder lines is easy to understand analytically in this model. In
hindsight, the phase structure of relativistic fermions is qualitatively
quite similar to that of static fermions. The absence of an exact
particle-hole symmetry leads to a distortion of the features found
in phase diagrams, but the overall behavior appears to be the same.
Both models exhibit eigenvalues of the mass matrix with negative real
parts for some regions of parameter space. Because the static fermion
case is tractable, we believe that this behavior is physical, indicating
regions where the phase of complex masses becomes sufficiently large.
There are, however, other possibilities. It is possible that the incorrect
solution is being used, or that the tree level approach fails when
this occurs. Another possibility, discussed below, is that we have
found regions where no equilibrium thermodynamic system exits. It
is striking that the case of relativisitic and static fermions have
liquid-gas phase transitions at tree level, while nonrelativisic fermions
and classical particles do not. It is not clear if this reflects some
fundamental feature of the physical systems being modeled, or some
limitation of the method used.

Many systems can be studied within the framework we have developed.
However, we are acutely aware that not much is known about the stability
of most of these systems. Sufficent conditions for the thermodynamic
stability of systems of classical particles were developed some time
ago by Fisher and Ruelle \cite{FisherRuelle1966}. For example, the
following conditions on the total potential $V=V_{2}-V_{1}$ are sufficient
for stability of a $d$-dimensional system: for some positive values
of $a_{1}$ and $a_{2}$, we have 
\begin{eqnarray}
r<a_{1} &  & V(r)\ge C/r^{d+\epsilon}\\
a_{1}<r<a_{2} &  & V(r)\ge-w\\
r>a_{2} &  & V(r)\ge-C'/r^{d+\epsilon'}
\end{eqnarray}
where $C$, $C'$, $w,$ $\epsilon$ and $\epsilon'$ are positive
constants. A system with attractive and repulsive Yukawa potentials
will satisfy thes conditions if $e>g$. To our knowledge, there are
no similar rigorous results for other systems. On physical grounds,
we expect that fermionic systems will be stable whenever the corresponding
classical system is stable. Because many of the current approaches
to the sign problem, including the one used here, rely on saddle points
in the complex plane of unknown stability, it would be very helpful
to know for which parameter values a given system is thermodynamically
stable.

\section*{Appendix}

In the common case where $F$ can be written as a function of $\Phi=\beta^{1/2}g\phi_{1}+i\beta^{1/2}e\phi_{2}$,
there is a formal equivalence between the partition function of the
effective field theory and a generalized Liouville sine-Gordon field
theory. This equivalence is a generalization of the equivalence of
the sine-Gordon model with a Coulomb gas \cite{doi:10.1063/1.1724281,Coleman:1974bu}.
The equivalence is proven by expanding $F\left(\Phi\right)$ in the
action in a power series in $z$ and integrating the resulting functional
integrals exactly at each order in the expansion. 
\begin{equation}
L_{3d}=\frac{1}{2}\left(\nabla\phi_{1}\right)^{2}+\frac{1}{2}m_{1}^{2}\phi_{1}^{2}+\frac{1}{2}\left(\nabla\phi_{2}\right)^{2}+\frac{1}{2}m_{2}^{2}\phi_{2}^{2}-F\left(\Phi\right).
\end{equation}
The function $F$ has a natural expansion of the form
\begin{equation}
F=\sum_{n}f_{n}e^{n\Phi}.
\end{equation}
Expansion in the $f_{n}$ leads to the interpretation of the partition
function as the grand canonical partition function as a gas with multiple
charges $n$ and fugacities $f_{n}$; some of the fugacities may be
negative. 

For simplicity, consider the case where only the $n=1$ term is nonzero.
Writing
\begin{equation}
F=\frac{z}{\lambda_{T}^{d}}e^{\Phi}
\end{equation}
we can expand the partition function in powers of $z$:, 
\begin{equation}
Z=\int\left[d\phi\right]e^{-S_{0}}\sum_{k=0}^{\infty}\frac{1}{k!}\left(\frac{z}{\lambda_{T}^{d}}\right)^{k}\int d^{d}x_{1}...d^{d}x_{k}\exp\left[\sum_{j=1}^{k}\Phi\left(x_{j}\right)\right]
\end{equation}
where
\begin{equation}
S_{0}=\int d^{d}x\left[\frac{1}{2}\left(\nabla\phi_{1}\right)^{2}+\frac{1}{2}m_{1}^{2}\phi_{1}^{2}+\frac{1}{2}\left(\nabla\phi_{2}\right)^{2}+\frac{1}{2}m_{2}^{2}\phi_{2}^{2}\right].
\end{equation}
Interchanging functional integration and summation and performing
the Gaussian functional integrals we have
\begin{equation}
Z=\sum_{k=0}^{\infty}\frac{1}{k!}\left(\frac{z}{\lambda_{T}^{d}}\right)^{k}\int d^{d}x_{1}...d^{d}x_{k}\exp\left\{ -\beta\sum_{k<l}\left[V_{2}\left(x_{k}-x_{l}\right)-V_{1}\left(x_{k}-x_{l}\right)\right]\right\} 
\end{equation}
where the Yukawa (screened Coulomb) potentials are determined by their
Fourier transforms
\begin{eqnarray}
\tilde{V}_{1}\left(q\right) & = & \frac{g_{1}^{2}}{q^{2}+m_{1}^{2}}\\
\tilde{V}_{2}\left(q\right) & = & \frac{g_{2}^{2}}{q^{2}+m_{2}^{2}}
\end{eqnarray}
The presence of terms in the expansion of $F$ with $n=1$ correspond
to higher charges in the generalized Coulomb gas representation.

\bibliographystyle{unsrtnat}
\bibliography{Liquid-Gas}

\end{document}